\documentclass[final,5p,times,twocolumn,number]{elsarticle}

\usepackage[T1]{fontenc}
\usepackage[utf8]{inputenc}

\makeatletter
\def\ps@pprintTitle{%
  \let\@oddhead\@empty
  \let\@evenhead\@empty
  \let\@oddfoot\@empty
  \let\@evenfoot\@oddfoot
}
\makeatother

\usepackage{booktabs}
\usepackage{array}
\usepackage{tabularx}
\usepackage{amssymb}
\usepackage{amsmath}
\usepackage{color,soul}
\usepackage{flushend}
\usepackage{stfloats}
\usepackage{listings}
\usepackage{xcolor}
\usepackage{hyperref}
\usepackage{moresize}
\usepackage{graphicx}
\graphicspath{{paper/}{./}}

\hypersetup{colorlinks=true,linkcolor=blue,urlcolor=blue,citecolor=blue}
\definecolor{codegreen}{rgb}{0,0.6,0}
\definecolor{codegray}{rgb}{0.5,0.5,0.5}
\definecolor{codepurple}{rgb}{0.58,0,0.82}
\definecolor{backcolour}{rgb}{0.95,0.95,0.92}

\lstdefinestyle{mystyle}{
  backgroundcolor=\color{backcolour}, commentstyle=\color{codegreen},
  keywordstyle=\color{magenta},
  numberstyle=\tiny\color{codegray},
  stringstyle=\color{codepurple},
  basicstyle=\ttfamily\footnotesize,
  breakatwhitespace=false,
  breaklines=true,
  captionpos=b,
  keepspaces=true,
  numbers=left,
  numbersep=5pt,
  showspaces=false,
  showstringspaces=false,
  showtabs=false,
  tabsize=2
}

\newcommand{\code}[1]{\texttt{#1}}
\newcommand{\invitem}[1]{\smallskip\noindent\textbf{#1:} }
\newcommand{\secref}[1]{\hyperref[#1]{Section~\ref*{#1}}}
\newcommand{\secsref}[2]{\hyperref[#1]{Sections~\ref*{#1}} and~\hyperref[#2]{\ref*{#2}}}
\newcommand{\tabref}[1]{\hyperref[#1]{Table~\ref*{#1}}}
\newcommand{\figref}[1]{\hyperref[#1]{Figure~\ref*{#1}}}
\makeatletter
\renewcommand\subsection{\@startsection{subsection}{2}{\z@}%
           {12\p@ \@plus 6\p@ \@minus 3\p@}%
           {3\p@ \@plus 6\p@ \@minus 3\p@}%
           {\normalfont\normalsize\bfseries}}
\makeatother

\begin{document}

\begin{frontmatter}

\title{PaxosLease Revisited: A Checked Model of Diskless Distributed Leases}

\author{M\'arton Trencs\'eni (\code{mtrencseni@gmail.com})}

\begin{abstract}
PaxosLease is a protocol by which a quorum of acceptors grants time-bounded exclusive ownership with no durable acceptor lease state and no disk write on the lease acquisition path.  This paper gives a precise, machine-checked statement of the protocol and of its standard use, electing a Multi-Paxos leader.  Formalizing and model checking the original protocol changes three rules of its acceptor: two are required for safety, the third allows shorter restart quarantines.  The protocol is formalized in TLA+ and checked by TLC, its timing arithmetic is proved in TLAPS, and an executable Python model demonstrates the distributed algorithm for human readers.
\end{abstract}

\end{frontmatter}

\section{Introduction}\label{sec:intro}

PaxosLease \cite{ref1} is a protocol by which a quorum of acceptors grants time-bounded exclusive ownership, diskless in the sense that acceptors do not write to disk, and proposers write to disk only once, at startup.  This paper gives a precise, machine-checked statement of the protocol and of its standard use, electing a Multi-Paxos leader that recovers once per epoch and then commits with single-round appends.

A lease is a lock whose ownership is bounded by time and expires automatically \cite{ref3}, so a holder that has crashed or become unreachable before giving up the lease does not block progress in the distributed system.  A process may act as owner only until its lease expires, and a process that granted the lease refuses conflicting grants until its own exclusion interval has expired.  Exclusion is the property that at any time\footnote{This paper implicitly assumes Newtonian relativity, where ``at any time'' has an absolute, frame-independent meaning.  A companion paper explores the modification of PaxosLease to special relativity with moving inertial observers.} at most one process has the authority to act as owner.  PaxosLease achieves exclusion with ballots and process identities carried in messages, local timers, and majority quorums.  Clocks need not be synchronized, but the intervals measured by different processes must have bounded error.

PaxosLease was described in 2012 \cite{ref1} and implemented in the open source systems Keyspace \cite{ref14} and ScalienDB \cite{ref15}.  Neither the description nor the implementations were model checked, and both contain subtle bugs, in the algorithm and in the C++ code.  This paper states the protocol precisely, checks it in TLA+, and changes three rules of its acceptor: an acceptor that accepts a request under a ballot rejects any later request under a lower ballot; it must never replace another owner's live lease, even under a higher ballot (\secref{sec:designspace}); and a restarted acceptor quarantines itself only for the proposer attempt duration (\secref{sec:safety}).  \secref{sec:designspace} and \ref{app:latetimer} discuss the bugs in the 2012 description, and \secref{sec:audit} and \ref{app:audit} those in the implementations, each supported by a TLA+ model or an executable witness.

\secref{sec:outline} describes the protocol in plain terms, \secref{sec:algorithm} states the system model and the algorithm, \secref{sec:safety} gives the safety argument and the quarantine bound, and \secref{sec:checked} summarizes the checked models.  \secref{sec:designspace} gives the safety violations of the original formulation, \secref{sec:audit} audits the historic implementations, and \secsref{sec:related}{sec:limitations} discuss related work and limitations.  The appendices give the leased-leader algorithm, the checked invariants, the executions with overwriting acceptors, the audit in full, the evidence inventory, and how the paper was written.

\section{How PaxosLease Works}\label{sec:outline}

PaxosLease grants one process a time-bounded, exclusive right to act as owner of a (potentially abstract) resource.  The idea is to borrow the structure of Paxos, its ballots, its two rounds and its majorities of acceptors, but for a value that is not a log entry: a statement of the form ``node 3 owns the lease, for the next seven seconds.''  A lease has a bounded validity interval, so an acceptor need remember it only until it expires, and can keep it in memory; PaxosLease acceptors do not write to disk.

\begin{figure}[t]
\centering
\includegraphics[width=\columnwidth]{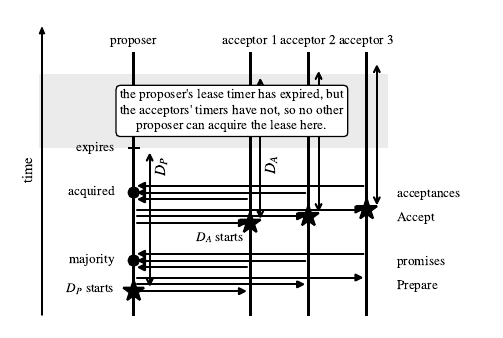}
\caption{\label{fig:acquisition}One acquisition in PaxosLease, as a message-sequence diagram.  The proposer's timer $D_P$ starts when \code{Prepare} is sent, each acceptor's timer $D_A$ starts when it accepts, and $D_P$ ends first.}
\end{figure}

An acquisition is two rounds.  A proposer picks a ballot number that nobody has used before and sends it to every acceptor in a \code{Prepare} message: the acceptors promise to reject any smaller ballot from now on, and respond whether they currently hold a live lease for another owner.  If a majority of acceptors answer, and none of them reports a live lease belonging to another owner, the proposer sends a second, \code{Accept} message\footnote{The 2012 paper \cite{ref1}, and the Keyspace \cite{ref14} and ScalienDB \cite{ref15} implementations built on it, call this second round \emph{propose}.  This paper uses \emph{accept}, the name the round carries in \cite{ref2}, and keeps \emph{propose} only where the text describes the original pseudocode.} asking the acceptors to record it as the owner for a fixed duration.  An acceptor that holds a live lease for another owner refuses to accept until its own timer has expired.  When a majority confirm, the proposer is the owner, and it informs the learners.  Figure~\ref*{fig:acquisition} shows the two rounds as a message-sequence diagram.

Any two majorities of the same set share at least one member, so two proposers cannot both collect a majority of acceptances for leases that overlap in time: the acceptor the two majorities share holds the first lease for at least as long as its owner may act on it, and refuses the second.  The first round lets a proposer find a live lease before it asks for acceptances it would not get, and the ballots let acceptors discard the messages of superseded attempts.

What Paxos does not have to reason about, and this protocol does, is time.  Two obligations follow.  First, an owner's authority must end no later than the acceptors' exclusion; if it ended later, a second proposer could acquire while the first still believes itself the owner.  Second, an acceptor that crashes loses its promises and its accepted lease, and if it answered again at once it could help a second proposer acquire a lease that is already held.  To address this, a restarted acceptor refuses all lease-layer messages for a fixed quarantine interval on restart.

A lease is renewed by running the same two rounds again, under a higher ballot, before the current one expires.  A lease may also be released early; this is an optimization that speeds up the next acquisition, but is never required for safety.

The standard use of PaxosLease is to elect a Multi-Paxos leader.  While a proposer holds the lease no other proposer can acquire it, so no other proposer runs the Paxos prepare round (Phase~1).  The holder runs Phase~1 once for the whole log and then commits each value with the accept round (Phase~2) alone: one round trip and one durable write per acceptor instead of two, on a fast path that competing ballots do not preempt.  The lease also lets the leader serve reads from its local state without a Paxos round, since no other process can commit writes while it holds the lease.  \ref{sec:paxos} states the leader lifecycle and the composition model that checks it.

\section{PaxosLease Algorithm}\label{sec:algorithm}

A reader who wants to implement PaxosLease should work from this definition, and not the original in \cite{ref1}.  We state the algorithm as numbered steps.  Every step names, in brackets, the action of \code{tla/spec/PaxosLease.tla} that formalizes it.  The TLA+ module, not this prose, is the normative statement of the PaxosLease algorithm.

\subsection{System Model}\label{sec:model}

The system contains a finite set of proposers and a finite set of acceptors.  A quorum is a set of acceptors such that any two quorums share at least one acceptor; specifically the paper uses majorities.

Processes are not Byzantine, but any process may crash.  A crashed acceptor loses its promised ballot and its accepted lease.  A proposer keeps one word on stable storage, a restart counter incremented at every start, which makes its ballots unique across restarts.

Messages may be delayed, reordered, duplicated, or lost, and a message sent to a process before it crashed may be delivered after it restarts.  The models also check the connection-oriented alternative, in which a crash discards the messages addressed to the crashed process (\ref{app:audit}).

Clocks are not synchronized, and messages carry no timestamps.  Each process measures elapsed time on its own clock, and the rate of every clock is within a factor $1 \pm \rho$ of real time, with $0 \leq \rho < 1$.

\subsection{State}

Proposer state is per attempt and volatile, with one durable exception.

\begin{itemize}\itemsep0.25em
\item the current ballot $b$, and the phase of the attempt (idle, preparing, accepting);
\item the pending deadline $t_d$ of the attempt, set when \code{Prepare} is broadcast (T1) and checked by P3 and P4;
\item the renewal base: the ballot of the lease this attempt renews, or none if the attempt is a fresh acquisition (P1, P2);
\item the identities of the acceptors that have responded in the current attempt, kept as a set rather than a count (P2);
\item when active, the ballot of the active instance, the authority deadline, and the quorum certificate that established the lease;
\item the restart counter, which is the one piece of proposer state written to stable storage (\secref{sec:model}).
\end{itemize}

Acceptor state is entirely volatile, which is what makes the protocol diskless.

\begin{itemize}\itemsep0.25em
\item the highest ballot promised since the last restart;
\item the accepted lease instance, if one exists, and the local deadline until which it excludes conflicting grants; the accepted lease is \emph{live} until that deadline passes on the acceptor's clock.
\end{itemize}

A lease instance is identified by $(owner, ballot)$, where the ballot is globally unique, also across proposer restarts.  The owner is a proposer incarnation: the proposer together with the restart counter its ballots carry (P1), so a proposer that restarts is a new owner.  This is required for safety after restarts.

\subsection{Durations}\label{sec:durations}

PaxosLease uses three fixed timeout durations, $D_P$, $D_A$ and $D_Q$.  Each is measured on the clock of the process that uses it, and $\rho$ bounds the rate error of every clock (\secref{sec:model}).

\begin{itemize}\itemsep0.25em
\item $D_P$, the proposer lease duration: an attempt whose \code{Prepare} is sent at time $t$ can make its proposer active only until $t + D_P$ (P1, T1).
\item $D_A$, the acceptor lease duration: an acceptor that accepts a lease refuses conflicting grants for $D_A$ time after the acceptance (A2).
\item $D_Q$, the restart quarantine: a restarted acceptor refuses every lease-layer message for $D_Q$ time (A4).
\end{itemize}

The timing rules T2 and T3 require $D_A$ and $D_Q$ to be at least $D_P \cdot (1+\rho)/(1-\rho)$.

\subsection{Proposer}

\invitem{P1 [\code{StartAcquire}]}Choose a fresh ballot $b$, never used by any proposer, including this proposer's own earlier incarnations.  A ballot is a triple $(counter, restart, node)$ with counter-major comparison: ballots compare lexicographically, with $counter$ most significant and $node$ as the final tiebreak.  The $counter$ increments per attempt, and $restart$ is durably stored and incremented at every process start, so no ballot is reused across proposer restarts.  Safety needs only this uniqueness, together with the $restart$ component, which identifies the incarnation for A2.  The counter does not order a restarted proposer's ballots above its earlier ones, since $(1, restart{+}1, node) < (100, restart, node)$; for liveness, a proposer raises its counter above any ballot it observes in responses, which requires rejections that carry the promised ballot (A1).  Record as the attempt's \emph{renewal base} the ballot of the lease this proposer currently holds, or none if this is a fresh acquisition; P2 tests self-owned reports against it.  Assign the pending deadline $t_d := now + D_P$ \emph{as stored state}, and broadcast \code{Prepare(b)}.

\invitem{P2 [\code{DeliverPromise}]}On a promise for the current $b$, record the responder's \emph{identity} and call the response \emph{open} if it reports no live lease, or, for a renewal attempt, if it reports exactly the lease being renewed (the renewal base captured at P1) while the proposer still holds that lease.  Quorums are sets of distinct acceptors (not just counts of messages, see \secref{sec:audit}, fourth finding).  A self-owned accepted lease under any other ballot may have been installed by an \code{Accept} request an abandoned attempt left in flight, and it blocks like a foreign lease.  With A2's acceptor-side rule the qualifier is not needed for exclusion (\secref{sec:safety}); it is needed with acceptors that overwrite (\ref{app:latetimer}).

\invitem{P3 [\code{SendAccept}]}When open promises from a quorum are recorded and $now < t_d$: broadcast \code{Accept(b, self)}.

\invitem{P4 [\code{DeliverAccepted}]}When \code{Accepted} responses for $b$ have arrived from a quorum of distinct acceptors and $now < t_d$, become active until $t_d$, record the quorum as the certificate, pass $t_d$ to its own learner, and broadcast \code{LearnChosen(self, t\_d - now)}, which carries the remaining duration; later responses under $b$ are discarded.

\invitem{P5 [\code{AbandonAttempt}, then P1]}An attempt that cannot proceed is abandoned, at the latest when $now \geq t_d$: the proposer returns to idle, alive and keeping whatever lease it already holds, and retries with a fresh, higher ballot from P1.  Messages of the abandoned attempt may still be in flight, and the ballot comparison in P2 and P4 discards every response belonging to it.  Retry does not advance the durable restart counter; only a crash and restart does.

\invitem{P6 [\code{Release}, optional]}To release, first stop acting as owner, then broadcast \code{Release}$(b_a)$, where $b_a$ is the ballot of the currently acquired lease.  An attempt in progress may be abandoned or left to complete; either is safe (\secref{sec:safety}).  Release is an availability optimization, not a safety requirement.

\smallskip
Messages may be lost, so in practice a proposer abandons an attempt and retries (P5) after a timeout, or as soon as the responses it has collected show that it cannot obtain a majority.

A renewal is a new acquisition by the current owner, under a new, higher ballot.  At P1 the attempt records the ballot of the lease it extends as its renewal base.

\subsection{Acceptor}

\invitem{A1 [\code{DeliverPrepare}]}On \code{Prepare(b)}: if $b \geq$ the promised ballot, set promised $:= b$ and reply with the currently live accepted lease, reporting an expired lease as no lease; otherwise reply rejecting, or not at all.  A rejection is a speed optimization: it lets a proposer learn early that it cannot obtain a majority.  So that proposers can raise their counters (P1), it should carry the promised ballot.  Safety does not depend on rejections: an acceptor that stays silent refuses just as effectively, and the proposer's own timeout ends the attempt either way.

\invitem{A2 [\code{DeliverAcceptReq}]}On \code{Accept(b, o)}: if $b \geq$ promised and the acceptor holds no live lease of an owner other than $o$, set promised $:= b$, record the lease $(o, b)$ with exclusion deadline $now + D_A$, and reply accepted; otherwise reject.  The second condition is the acceptor-side rule of \secref{sec:designspace}: without it, a renewal across a single acceptor restart gives two owners, and so does early release.  Rejecting lower ballots after an accept is also a change from \cite{ref1}, whose acceptor compares an \code{Accept} request only against the ballots of earlier \code{Prepare} requests: the first case of \secref{sec:safety} uses it, and without it a delayed \code{Accept} request and \code{Release} of the owner's previous acquisition can clear an acceptor of the owner's current lease, with no crash.  Owners are compared as proposer incarnations.  A restarted proposer's ballots may be lower than its previous incarnation's, and a delayed \code{Accept} request of the previous incarnation must not replace the new incarnation's lease (\secref{sec:safety}).

\invitem{A3 [\code{DeliverRelease}]}On \code{Release} from $o$ with ballot $b$: clear the accepted lease only if it matches the message's $(o, b)$ exactly.  Matching on the owner $o$ alone leads to a safety violation (\tabref{tab:variants}, 36-state trace).  Exact matching is necessary but not sufficient: without A2's refusal rule, a stale \code{Accept} can reinstall a released instance over another owner's live lease, and the stale \code{Release} then clears it, exclusion deadline included (\secref{sec:designspace}).

\invitem{A4 [\code{RestartAcceptor}]}On restart after losing volatile state: refuse every lease-layer message, prepare, accept, and release alike, until $now + D_Q$.

\subsection{Learner}

\invitem{L1}Activation is made application-visible through learners.  The owner's own learner installs the absolute deadline $t_d$, and only the owner acts on the lease.  A remote learner installs the remaining duration from the message's arrival, minus an assumed bound on message delivery time, so its view of the lease is only a hint (\secref{sec:audit}, sixth finding): it may report a lease that has already expired or passed to another proposer.  This is safe as long as the view is used only as an optimization, for example to keep the local proposer from starting an acquisition while a lease is reported.  In the worst case a proposer waits longer than necessary, or starts an attempt that the acceptors refuse, and the system is without a leader for longer time between leases.

\subsection{Timing Rules}

This section enumerates the PaxosLease timing constraints as named rules for proposer attempts, lease containment, and acceptor restart quarantine.

\invitem{T1 (proposer lease timeout)}The pending deadline $t_d$ is assigned when \code{Prepare} is broadcast (P1) and checked in the transitions that use the attempt's evidence (P3, P4).  Safety requires only that the timer start no later than the first acceptor accepts the lease (\secref{sec:safety}).  Starting it when \code{Prepare} is sent is needed only for the overwriting acceptors of \cite{ref1}, for which a later start is unsafe (\ref{app:latetimer}).

\invitem{T2 (acceptor lease timeout)}An owner's authority must end no later than the exclusion interval of every acceptor in its certificate.  With equal clock rates $D_P \leq D_A$ suffices, since the proposer's timer starts no later than any acceptance (T1).  With rate bounds $r_{min} = 1-\rho$ and $r_{max} = 1+\rho$ the condition is
\[
  D_P \cdot r_{max} \leq D_A \cdot r_{min},
\]
which covers a proposer clock running slow and an acceptor clock running fast; the TLAPS obligation \code{DriftContainmentArithmetic} proves the arithmetic.

\invitem{T3 (acceptor restart quarantine)}With equal clock rates, $D_Q \geq D_P$ suffices.  With the rate bounds of T2 the condition is
\[
  D_P \cdot r_{max} \leq D_Q \cdot r_{min}.
\]
Neither the acceptor lease duration $D_A$ nor the maximal lease time appears in it; \secref{sec:safety} shows that it suffices and, by checking in two-acceptor acquisition configurations, that it is tight even when $D_P < D_A$.

\section{Safety Argument}\label{sec:safety}

\medskip
\noindent\textbf{Theorem.} At any time, at most one proposer is active.
\medskip

The theorem is the invariant \code{LeaseExclusivity} of \code{tla/spec/PaxosLease.tla}; every completed configuration of \code{PaxosLeaseChecked.tla}, the module that assumes T3 and A2's rule, passes it.  The following argument assumes $\rho = 0$ for simplicity; it is not a mechanized proof.

\emph{Argument.}  Suppose that two different proposers $p$ and $q$ are both active at time $t_v$, violating safety.  Each is active under a lease instance, $(p, b_p)$ and $(q, b_q)$, whose \code{Accepted} responses came from a quorum, its certificate.  Any two quorums intersect, so some acceptor $a$ is in both certificates and accepted both instances.  Name the proposers so that $a$ accepted $(p, b_p)$ first, at time $t_p$, and $(q, b_q)$ later, at time $t_q$.  Since $q$ became active only after $a$'s response, $t_q \leq t_v$.

By timer containment, $a$'s exclusion deadline for $(p, b_p)$ is no earlier than the end of $p$'s authority under that instance, which lies after $t_v$, so on $a$'s own clock the lease was still live at $t_q$.  A2 then refuses $(q, b_q)$ unless $a$ lost its accepted lease of $p$ between $t_p$ and $t_q$.  Only an \code{Accept} request, a release, and a crash change an acceptor's accepted lease before it expires.
\begin{itemize}\itemsep0.25em
\item An \code{Accept} request replaces a live accepted lease only when it names the same owner, the same incarnation of $p$ (A2), and only under a ballot no lower than $a$'s promised ballot, which is at least $b_p$ from $t_p$ on.  The replacement is therefore an accepted lease of that incarnation under a ballot of at least $b_p$, with a later exclusion deadline, and never one of $q$, so $a$ still holds a live accepted lease of $p$ at $t_q$ and refuses $(q, b_q)$.
\item A release clears an accepted lease only when it names that exact instance (A3), and a proposer releases only the instance it is active under, after it has stopped acting under it (P6).  A release of $(p, b_p)$ before $t_v$ contradicts $p$ being active under $(p, b_p)$ at $t_v$, since an instance activates at most once (P1, P4).  A release of a lower instance of the same incarnation matches nothing, since from $t_p$ on $a$'s accepted lease carries a ballot of at least $b_p$ (first case).  A release of a higher instance contradicts it as well: $p$ was active under that higher instance when it released it, before $t_v$, and an incarnation's active ballot never decreases, so it cannot be active under $(p, b_p)$ at $t_v$.
\item If $a$ crashed between $t_p$ and $t_q$, it could not accept $(q, b_q)$ until $D_Q \geq D_P$ after its restart (A4, T3).  The pending deadline of $(p, b_p)$ was assigned no later than $a$ accepted it (T1), so $p$'s authority under that instance ended at most $D_P$ after $t_p$, no later than $t_q$, which contradicts $p$ being active at $t_v$.
\end{itemize}
Every case contradicts the assumption, so at most one proposer is active at any time.  The same argument holds with clock-rate error $\rho$ and the rate-adjusted forms of the inequalities T2 and T3.

The crash case shows that $D_Q \geq D_P$ suffices: the acceptor exclusion duration $D_A$ does not appear, and with A2's rule a forgotten promise imposes no bound of its own.  The bound is tight in the checked model: at two separated duration pairs, a quarantine of $D_P$ passes exhaustively and one of $D_P - 1$ yields a two-owner trace (\tabref{tab:configs}), in two-proposer, two-acceptor acquisition configurations.

The argument uses neither \code{Prepare} responses nor P2's renewal qualifier, so with A2's rule these only make acquisition efficient; with overwriting acceptors they are needed for safety (\ref{app:latetimer}).

This is a safety argument only.  Liveness requires eventual message delivery, sufficiently accurate timers, and a period during which a quorum of acceptors stays available; as in Paxos, two proposers can starve one another with ever higher ballots, which implementations prevent with randomized backoff \cite{ref1}.

\smallskip
\noindent\textbf{A2 compares owners as incarnations.}  A proposer's attempt counter is volatile, so a proposer that restarts starts it again under a higher restart component, and its new ballots may be lower than its old ones (P1).  If A2 compared owners by proposer alone, a stale message of the previous incarnation could pass as the current owner's.  Proposer $p_1$ acquires under ballot~2 through $a_1$ and $a_2$ and releases, with its \code{Accept} request and its \code{Release} to $a_3$ still in flight.  It then crashes and restarts, $a_2$ crashes and restarts into the full quarantine, and $p_1$ acquires again under the lower ballot~1, through $a_2$ and $a_3$.  The stale \code{Accept} request of ballot~2 reaches $a_3$; its ballot is higher and its proposer is the same, so a comparison by proposer alone replaces $p_1$'s live lease with the released instance, and the stale \code{Release} then clears it.  Proposer $p_2$ acquires through $a_1$ and $a_3$ while $p_1$ is active.  TLC finds the execution in a 43-state trace on the variant \code{NodeOwner.tla} (\tabref{tab:variants}), and the same directed search on the specification, which compares incarnations, exhausts its space with no violation.  The reference model reproduces an execution of the same shape (\code{release-across-proposer-restart}).  The clause is needed because of the counter-major layout of P1, kept from \cite{ref1}.  With the restart counter above the attempt counter, a proposer's ballots would increase across its restarts and A2's ballot guard alone would reject the previous incarnation's stale \code{Accept} request, but a proposer that had restarted fewer times could then outbid another only by raising its own restart counter, a durable write.

\section{Checked Models and Tests}\label{sec:checked}

Unlike the 2012 formulation, the protocol of \secref{sec:algorithm} comes with a formal TLA+ specification that TLC checks exhaustively in finite configurations, with the timing arithmetic proved in TLAPS, and with an executable reference model that encodes the same rules a second time.  The \href{https://github.com/mtrencseni/paxoslease-revisited}{accompanying GitHub repository} (\ref{app:artifact}) contains:
\begin{itemize}\itemsep0.1em
\item the TLA+ specification and its weakened variants;
\item the TLAPS proof module;
\item a Python reference model of the lease protocol and of its composition with Multi-Paxos;
\item eighteen structured witnesses that keep every known failure executable, and the tests;
\item a single-file runnable demonstration;
\item the recorded output of every run.
\end{itemize}

TLC checks \code{LeaseExclusivity} together with the supporting invariants of \ref{app:invariants}, and \tabref{tab:evidence} maps each rule of \secref{sec:algorithm} to the model, variant, witness or audit that checks it.  Each variant weakens exactly one rule, two of them on the overwriting acceptor of \cite{ref1} (\tabref{tab:variants}), and is generated from the specification by a script, and TLC finds a two-owner execution in each (\tabref{tab:variants}).  A second TLA+ model projects the protocol the historic implementations ship (\ref{app:audit}).

The passing configurations cover acquisition races, acceptor crash and restart, renewal, release, abandoned attempts and a redelivering transport, separately and in combination; renewal and release are combined only with a single proposer.  The passing checks are exhaustive only within small configurations: at most two proposers, three acceptors, three ballots, one restart per proposer, a time horizon of three units, no message loss outside the directed renewal search, and abandoned attempts (P5) only in the Retry rows, with one duration triple $(D_P, D_A, D_Q)$ per configuration (\tabref{tab:configs}).  Even within these bounds the recorded runs explore more than five billion distinct states, and TLC generates more than thirty billion states along the way; the three largest completed runs, Retry, Renew+Retry, and the stale-owner check with A2's rule (\ref{app:latetimer}), each exceed two billion distinct states.

\section{Safety Violations in the Original Formulation}\label{sec:designspace}

The 2012 PaxosLease protocol existed as a paper and open source implementations for over a decade with no machine-checked model.  The 2012 paper \cite{ref1} does not state the start of the proposer's timer consistently.  Its figure is itself ambiguous: its row of event labels starts the timer before the \code{Prepare} requests, the rule this paper specifies as T1, while the marker on the proposer's line starts it after the prepare responses; its pseudocode states the quorum-receipt rule, which \ref{app:latetimer} shows unsafe with the overwriting acceptors that \cite{ref1} specifies; and its prose is consistent with both.

In \cite{ref1}, an acceptor accepts every \code{Accept} request with a high enough ballot, even when it already holds a live lease for another owner: the old lease is simply overwritten.  Keyspace and ScalienDB do the same.  A2 changes this, and an acceptor refuses to replace a live lease of another owner.  The two executions below show why the change is needed: with overwriting acceptors, each gives two owners at once, even with the timer rule T1 and the renewal check of P2 in place.  With A2's rule, T1's timer placement and P2's renewal check are no longer needed for safety; \ref{app:latetimer} shows what goes wrong without them when acceptors overwrite.

\smallskip
\noindent\textbf{Renewal across an acceptor restart.}  With three acceptors, one acceptor restart lets a renewal produce two owners.  The restarted acceptor has forgotten its promise to one proposer, whose lease rests on an accepted lease it overwrote at another acceptor; after its quarantine it accepts the other proposer's delayed \code{Accept} request and completes that proposer's majority.  TLC finds the execution in a 39-state trace, and the same search with A2's rule finds none (\ref{app:latetimer}).

\smallskip
\noindent\textbf{Early release across an acceptor restart.}  A delayed \code{Accept} request puts a released lease back at an acceptor, over another owner's live lease, and the delayed \code{Release} then clears it, so a third acquisition succeeds while that owner is still active.  No quarantine length prevents this; TLC finds it in a 40-state trace, and with A2's rule the search finds none (\ref{app:latetimer}).

\begin{table*}[tbp]
\centering
\small
\renewcommand{\arraystretch}{1.35}
\begin{tabular*}{\textwidth}{@{\extracolsep{\fill}}llll}
\toprule
\textbf{Variant} & \textbf{Weakened rule} & \textbf{Violation} & \textbf{Trace} \\
\midrule[\heavyrulewidth]
\code{LateTimer} & timer starts at prepare-quorum receipt & \code{LeaseExclusivity} & 27 states \\
\hline
\code{OwnerOnlyRelease} & release matches owner, ignores ballot & \code{LeaseExclusivity} & 36 states \\
\hline
\code{ScalarQuorumCounting} & quorum counted by message, redelivering transport & \code{LeaseExclusivity} & 18 states \\
\hline
\code{StaleOwnerOpen} & own lease open without the renewal qualifier, retry enabled & \code{LeaseExclusivity} & 36 states \\
\hline
\code{NodeOwner} & A2 compares owners by proposer, not incarnation & \code{LeaseExclusivity} & 43 states \\
\bottomrule
\end{tabular*}
\caption{\label{tab:variants}Counterexample variants: full copies of the specification with exactly one rule weakened, regenerated from it and compared against the checked-in files by \code{make variant-check}.  The target \code{make counterexamples} requires TLC to find the violations of \code{LateTimer}, \code{OwnerOnlyRelease}, \code{ScalarQuorumCounting} and \code{NodeOwner} on every run, the last through a directed search whose state constraint only prunes behaviors; the \code{StaleOwnerOpen} search is multi-hour and is recorded evidence, checked against the cited trace length by \code{make paper-claims}.  The \code{LateTimer} and \code{StaleOwnerOpen} configurations use the full quarantine bound and overwriting acceptors, the rule of \cite{ref1}; with A2's rule both pass.  Their acceptors still reject lower ballots after an accept, which only removes behaviors, so the violations are also executions of the acceptor of \cite{ref1}.}
\end{table*}

\section{Audit of the Historic Implementations}\label{sec:audit}

PaxosLease was implemented by this author in two open source systems, Keyspace \cite{ref14} and later ScalienDB \cite{ref15}.  This audit examines the final commits of both repositories against the checked model and against a checked projection of the shipped protocol, and makes eight findings, which \ref{app:audit} gives in full with the source-level evidence.

\begin{enumerate}\itemsep0.25em
\item Both systems implement the quorum-receipt timer rule of the pseudocode, which is unsafe with their overwriting acceptors, and an unrelated retry timeout of $R = 2$ seconds accidentally prevents the late-timer executions.
\item That prevention requires $D_Q \geq \max(D_P, R)$ under timely timeout dispatch, in acquisition configurations without renewal, and the shipped constants satisfy it.
\item The event loop can run an expired attempt's handlers before the retry timeout fires: TLC finds a two-owner execution under the shipped constants, also when only the quorum-intersection acceptor crashes.
\item Quorums are counted by message rather than by acceptor identity, which is safe only because their TCP transport does not deliver a response twice.
\item Every lease timer reads the wall clock, and neither system defends both of the directions in which a wall clock is unsafe.
\item Remote learners start the lease countdown at message arrival, which is safe only under an undocumented 500 msec delivery bound.
\item A proposer treats any lease it owns as an open response; with the shipped retry timeout, overwriting acceptors, and acceptors that crash independently of the proposers, TLC finds a two-owner execution with timely dispatch and no clock error.
\item The activation guard reads the clock twice, and an unsigned subtraction can wrap so that a proposer activates a lease that has already expired.
\end{enumerate}

In summary, both Keyspace and ScalienDB contain bugs through which two proposers can hold the lease at the same time: TLC finds such executions in the checked model of the shipped protocol (third and seventh findings), and wall-clock steps and the unsigned subtraction give further paths to an owner acting past its lease (fifth and eighth findings).

\section{Related Work}\label{sec:related}

Leases as time-bounded ownership originate with Gray and Cheriton \cite{ref3}.  Chubby \cite{ref6} made the lease-plus-Paxos architecture standard practice at scale, and the engineering account of \cite{ref7} describes how much of the difficulty lies outside the core algorithm.  FaTLease \cite{ref8} solves the same lease-negotiation problem as PaxosLease, but it runs Paxos instances for the lease commands and assumes clock synchrony, while PaxosLease was designed to remove both \cite{ref1}.  The closest relative is Flease \cite{ref10}, from the same group as FaTLease: decentralized lease coordination without stable storage for the lease state, built on a round-based register.

Raft \cite{ref4} structures leadership through terms and elections, but its safety, like that of Paxos, does not depend on real-time exclusivity.  Production Raft systems that serve reads from the leader without a log round trip add a leader lease, and they inherit exactly the containment and clock-rate obligations of T2.  Paxos Quorum Leases \cite{ref12} let a quorum of replicas serve local reads; they are a read-performance mechanism layered on Paxos rather than a diskless leadership protocol, but rest on the same containment argument.

Chand, Liu, and Stoller \cite{ref5} give a full TLAPS proof of Multi-Paxos, and Grove \cite{ref11} is the modern benchmark for mechanized lease reasoning, verifying time-based leases together with crash recovery, reconfiguration, thread-level concurrency, and unreliable networks.  Lamport's Paxos papers \cite{ref2,ref9} set the specification style, and the trace-validation work of Cirstea et al.\ \cite{ref13} shows how the remaining gap between such models and an implementation can be closed mechanically.

\section{Limitations}\label{sec:limitations}

The main limitations of this work are the following.

\begin{enumerate}\itemsep0.25em
\item PaxosLease assumes non-Byzantine processes that follow the protocol.
\item Only majority quorum systems\footnote{Alternatives include weighted majorities, grid quorums (a row plus a column), hierarchical majorities of majorities, and flexible quorums with different families for the two rounds.} were checked; the argument of \secref{sec:safety} uses quorum intersection.
\item The acceptor set is fixed.  Reconfiguration requires preserving quorum intersection across configurations, or else waiting until leases from the old configuration can no longer matter.
\item The TLC runs are finite-state checks within the bounds of \secref{sec:checked}, with a cap on messages in flight; every configuration with $D_P < D_A$ has two acceptors and neither renewal nor release.
\item With two proposers, crash and three acceptors, renewal and release are checked exhaustively only by the directed searches.  The undirected searches were stopped without a violation after 3.9 and 4.5 billion distinct states (\tabref{tab:configs}), with their queues still growing.
\item The clock model assumes monotonic elapsed-time measurement with bounded rate error.  A suspended, paused or live-migrated process can resume after its lease has expired and go on acting as owner.
\item Only the timing arithmetic is mechanically proved.  The exclusivity theorem rests on three things together, none of them a proof of it: the argument of \secref{sec:safety}, those arithmetic lemmas, and finite model checking.
\item The audit of \secref{sec:audit} is a reading of the sources against the model, backed by a checked projection and an executable abstraction of the node; we did not run the described executions against the shipped binaries, and the claims about margins assume the shipped compile-time constants.
\end{enumerate}

\section{Lessons}\label{sec:lessons}

Writing this paper taught the author four lessons that may apply to distributed algorithm research and practice beyond the scope of this paper.  The first three follow the steps of the development process, from specification to implementation; the fourth concerns automating them.

\begin{enumerate}\itemsep0.25em

\item \textbf{Formal specification and checks.}  Lamport's standing advice\footnote{In December 2009 the author described PaxosLease in an email to Leslie Lamport, who replied that he did not have time to look at the algorithm and suggested coding it in TLA+ and using the model checker to debug it.  This paper is that suggestion, carried out seventeen years later.  He was right on both counts: the model checker was the proper instrument, and the algorithm did need debugging.} is to write the algorithm in TLA+ and let the model checker debug it.

\item \textbf{Code derived from checked specifications.}  Production source code should be written from the checked TLA+ specification, not from pseudocode or prose without tool checks.  Both audited systems were written from the pseudocode of \cite{ref1}, which states the timer rule in its unsafe form.

\item \textbf{Code re-encoded as a specification and checked.}  Software adds networking and messaging code, event handling and control flow, concrete data types for the algorithm's variables, and other implementation detail, and each of these is an opportunity to introduce a bug into an otherwise correct algorithm.  By translating the software implementation back to a specification and then checking it, the risk of shipping unsafe code to production is reduced.

\item \textbf{Automation with frontier language models.}  The time required to write and maintain a specification, its configurations, its weakened variants and its harness has discouraged engineers from using formal methods.  Modern frontier models automate away much of this work.  However, extensive human guidance and review are required throughout the process (\ref{sec:llm}).

\end{enumerate}

\section{Conclusion}\label{sec:conclusion}

PaxosLease grants time-bounded exclusive ownership from a quorum of acceptors that keep no durable lease state.  Exclusion follows from quorum certificates whose acceptor exclusions outlast the proposer's authority, from acceptors that never replace another owner's live lease, where a restarted proposer is another owner, from release that clears only the exact instance named, and from a quarantine of one attempt duration after an acceptor restart.

\flushcolsend
\clearpage
\appendix
\gdef\thetable{\arabic{table}}
\footnotesize

\section{Using PaxosLease in Multi-Paxos}\label{sec:paxos}

PaxosLease is not Paxos, and the two should not be confused.  Paxos solves agreement for log slots, choosing at most one value in each slot \cite{ref2}.  Unlike the lease layer, its acceptors are durable: a promise and an accepted value must each reach stable storage before the acceptor replies, so committing one value in plain Paxos costs two round trips and two durable writes at every acceptor.

The leader optimization removes half of that.  A proposer runs Phase~1 once under a single ballot for the whole log rather than for one slot, and the promises it collects reserve that ballot for every slot it will later fill.  From then on it appends by running Phase~2 alone, so each new value costs one round trip and one durable write per acceptor instead of two.  Phase~1 is amortized, not skipped, and the reservation lasts until another proposer's own Phase~1 wins a quorum of promises under a higher ballot.

This is what the lease contributes.  A proposer holding a PaxosLease renews it, and may go on renewing it indefinitely: until it crashes, until the network prevents a renewal, or until the application hands leadership elsewhere.  While it holds the lease no other proposer can acquire the lease, and hence no other proposer will run a Prepare phase, so, unless a \code{Prepare} delayed from an earlier epoch arrives, the single reservation stays valid across arbitrarily many appends and the two round trips of the initial full round are amortized over all of them.  Driven to the limit, the amortized cost of committing a value is one round trip and one durable write at each acceptor, which is what makes leased Multi-Paxos a practical way to implement a replicated log.

What the lease does not do is replace Phase~1.  A proposer that has just acquired the lease does not know which values earlier leaders may already have had accepted, or chosen, and it must run a full first round to discover them and finish them before proposing anything of its own.

The standard use of the lease is this classic Multi-Paxos leader optimization; Keyspace and ScalienDB implement a per-slot form of it, described below.  We state the leader lifecycle precisely, in the form of \secref{sec:algorithm}: each step except M3, which the composition model abstracts, names its counterpart in the composition model (\code{tla/spec/PaxosLeasePaxos.tla}).  That model abstracts the lease layer to its conclusion: its \code{AcquireLease} action is enabled only while no proposer is active, so lease exclusivity is a premise there, and its invariants check the lifecycle above the lease; the Python composition runs the lease model underneath.

\invitem{M1 [\code{AcquireLease}]}Win the PaxosLease of \secref{sec:algorithm}.  This starts a fresh \emph{leadership epoch}, the interval over which one proposer holds the lease continuously, across any number of renewals, and it \emph{clears} any recovery standing from an earlier epoch.

\invitem{M2 [\code{StartRecovery}/\code{FinishRecovery}]}Run a log-wide Paxos Phase~1 under a fresh ballot.  In the Multi-Paxos organization assumed here, a promise quantifies over all instances rather than one \cite{ref2,ref9}, so the single round both reserves the ballot for the entire log and discovers the accepted values that constrain recovery.

\invitem{M3 [repair]}For every slot that Phase~1 reports constrained, select the value attached to the highest-ballot \code{Accepted} response for that slot, re-propose it under the reserved ballot, and learn the results.  Phase~1 alone is not cleanup: a value that is discovered but not re-proposed leaves recovery to be finished by whatever proposal next touches the slot.

\invitem{M4 [\code{BecomeReady}]}Only now is the leader ready.  Readiness is per epoch: the invariant \code{ReadyImpliesRecovered} checks that a re-elected leader repeats M2 rather than relying on a previous epoch's recovery; M3 is exercised by the Python composition.

\invitem{M5 [\code{AdmitOrApply}]}Fast mode.  Each client command is placed, unaltered, into a new slot with a single accept round under the reserved ballot and no per-instance \code{Prepare}.

Keyspace and ScalienDB organize Phase~1 differently.  Their Paxos acceptors reset the promise for every slot, and the leader skips \code{Prepare} on a new slot after committing the previous one within the same lease epoch, so no promise stands behind the skipped \code{Prepare}.  The agreement argument below therefore does not carry over to them, and this paper does not examine their fast path.

M1 relies on the exclusion theorem that \secref{sec:safety} establishes for the lease layer; none of M2 through M5 does.  Paxos agreement does not depend on lease exclusivity.  Acceptors reject lower-ballot accepts regardless of any clock, so even two processes that simultaneously believe themselves leader cannot choose conflicting values, and skipping \code{Prepare} in M5 is sound because M2 already promised the whole log.  The lease adds no safety to this and is needed for none of it.  What the lease adds is efficiency and liveness, since with at most one real-time leader the fast path is not repeatedly preempted by dueling ballots, and local admission and lease-guarded reads become possible.  One caution on reads: the lease alone does not make a local read current.  A leader serving reads locally needs the lease, this epoch's completed recovery (M2, M3), and an applied state machine covering the relevant prefix; ownership without recovery reads stale state.  Conversely, the fast mode imposes only one obligation on the leader: no admission before the leader's own epoch has completed M3.

The PaxosLease layer and the Paxos layer have a one-way interface, which is what makes them independently reviewable.  The PaxosLease layer determines which Paxos proposer may run M2 through M5, and when it may begin.  No Paxos transition reads PaxosLease state: a Paxos acceptor deciding whether to promise or to accept consults its own promised ballot and its own accepted values, and nothing else, and it does not know whether the Paxos proposer addressing it holds the lease.  No Paxos transition reads a clock: deadlines, durations, drift bounds and the quarantine all live in the PaxosLease layer, and agreement on the log is a function of the message history alone.  The optimization therefore changes only the Paxos proposer, which skips the prepare phase while the lease condition holds; the Paxos acceptor and the Paxos learner are unmodified.

\section{Checked Invariants}\label{app:invariants}

For P2's renewal clause, the TLA+ module stores no renewal base.  It tests at each response that the reported ballot is the proposer's active ballot, which within the module changes only through the attempt's own activation, so the test is equivalent to the stored base of the executable models.

The model has a variable \code{now}, and executable guards such as $now < pendingDeadline$ read it directly.  Since this can look like an assumption of synchronized clocks, we state precisely what it means.  The variable \code{now} is the common real-time axis on which safety is asserted, and each local timer is a duration that has been conservatively translated onto that axis using the rate bounds of T2.  When a process in the model compares \code{now} with its own deadline, it represents an implementation reading its own monotonic elapsed-time counter, never another process's clock.  Similarly, the certificate variables \code{cert} and \code{certDeadline} that appear in the invariants are history variables.  They record, for the benefit of the checked predicates, which acceptors granted the current lease and the exclusion deadlines those acceptors reported.  The protocol itself never compares a deadline measured on one clock with a deadline measured on another.

\subsection{Lease Invariants}

\invitem{\code{TypeOK}}Every variable remains in its declared finite domain: times are bounded, messages have one of the modeled shapes, acceptor state maps acceptors to lease records, proposer phases are among the modeled phases, and response sets are sets of acceptor identities.  This is the well-formedness condition needed before the other predicates have their intended meaning.

\invitem{\code{AcceptedCoherence}}An acceptor's accepted value is either \code{NoLease} or a lease whose owner, ballot, and deadline lie in the modeled domains.  This rules out malformed accepted state and ensures that later \code{Prepare} responses can be interpreted as lease reports.

\invitem{\code{ActiveImpliesUnexpired}}Every active proposer has a local deadline strictly greater than the current model time.  This captures the rule that a proposer stops acting when its own lease interval expires.

\invitem{\code{LeaseExclusivity}}The set of active proposers has cardinality at most one.  This is the main safety property.  It is checked directly by TLC, checked by the Python simulator after every operation, and violated on purpose by every counterexample variant.

\invitem{\code{ActivationHasQuorum}}An active proposer has a quorum certificate.  A proposer cannot become active merely because it sent requests or received a non-quorum set of responses.

\invitem{\code{TimerContainment}}For every acceptor in an active proposer's certificate, the proposer's deadline comes no later than the acceptor deadline recorded in the certificate.  The certificate variables are history variables (defined at the start of this appendix), so the invariant is a statement checked by the model, not a comparison performed by the protocol.

\invitem{\code{QuarantinePreventsParticipation}}An acceptor in quarantine has no promised ballot and no accepted lease.  Together with the transition rules, this means that a restarted acceptor cannot vote while forgotten lease state could still matter.

\invitem{\code{ActiveBallotWellFormed}}An active proposer's active ballot is one of the modeled ballots.  This is a bookkeeping invariant for the certificate machinery.  The substantive renewal property, that a failed renewal does not extend authority, is a transition rule exercised by the renew/release configuration and by the scenario tests.

\subsection{PaxosLease+Paxos Invariants}

\invitem{\code{ChosenValueWellFormed}}The abstract composition model represents each slot by a single chosen value or by \code{NoValue}, and the transition relation assigns a chosen value only when the slot is empty.  Agreement is therefore enforced by construction in this abstraction, the named invariant checks domain membership, and the substantive checks are the admission invariants below.  We state this explicitly to avoid overclaiming, since the composition model checks the lease/Paxos boundary, not Paxos itself.

\invitem{\code{ReadyImpliesActive}}A ready leader holds a live lease.  This is the admission side of the composition, since a process without lease authority cannot be ready to accept client log work.

\invitem{\code{ReadyLeaderUniqueness}}At most one proposer is in the ready state.  This is stronger than Paxos requires for agreement.  In the composition model it follows from the single-holder premise of \code{AcquireLease} together with \code{ReadyImpliesActive}, so it records the property the lease layer is assumed to supply.

\invitem{\code{RecoveryPrecedesAdmission}}A proposer cannot apply or admit log work before it has completed recovery.  This is the modeled boundary between the lease protocol and Paxos recovery.

\invitem{\code{ReadyImpliesRecovered}}A ready leader has completed recovery within its current lease epoch, since acquiring a lease clears the recovery mark.  This fences the fast path per epoch: no leader appends on the strength of a previous epoch's recovery.

\invitem{\code{AppliedImpliesChosen}}Every slot applied by any proposer has a chosen value, so the leader-admission abstraction cannot invent applied log entries that Paxos has not chosen.  \textbf{Log Prefix Consistency} is the two-proposer restatement of the same fact, implied by it and retained as a separate named check for readability.

\subsection{TLAPS Obligations}

\invitem{\code{TimerContainmentArithmetic}}If the proposer starts its timer no later than each certificate acceptor starts its exclusion, and its duration is no longer, then the proposer's authority interval is contained in the acceptor's exclusion interval.

\invitem{\code{QuarantineCoversForgottenFacts}}An attempt whose \code{Prepare} was sent no later than the crash is unusable $D_P$ later, so a quarantine of at least $D_P$ from the restart outlasts it.  The argument of \secref{sec:safety} supplies the premise, that everything an acceptor can forget belongs to such an attempt; quarantine at least $D_P$ therefore outlasts every attempt whose \code{Prepare} preceded the crash, which with A2's rule is every attempt a forgotten fact can support.  The obligation has no analogue under the step-3 rule alone; with a retry bound the analogue is the next obligation.

\invitem{\code{ImplQuarantineCoversForgottenFacts}}The analogue for the implemented protocol under timely timeout dispatch: a forgotten promise anchors at a \code{Prepare} and is unusable $R$ later, a forgotten accepted lease anchors at a prepare-quorum receipt and is unusable $D_P$ later, and both anchors precede the crash, so quarantine at least $\max(D_P, R)$ covers both cases.

\invitem{\code{CapMonotone}}The deadline cap that keeps the finite model bounded is monotone, so capping preserves containment.

\invitem{\code{QuarantineBoundTight}}Any quarantine strictly below the proposer duration leaves an elapsed interval inside a forgotten attempt's lifetime but outside quarantine.  The two-owner traces of \secref{sec:safety} lie in that gap.

\invitem{\code{DriftContainmentArithmetic}}If the proposer's clock runs at rate at least $r_{min}$, the acceptor's at most $r_{max}$, and the durations satisfy $D_P \cdot r_{max} \leq D_A \cdot r_{min}$, then real-time containment holds.  This is the inequality of T2, and the same arithmetic with the quarantine in place of $D_A$ gives the real-time quarantine bound.  The lemma is stated over natural numbers; rational rates and durations reduce to it by scaling to a common unit.

\section{Executions with Overwriting Acceptors}\label{app:latetimer}

This appendix gives the executions that overwriting acceptors admit: the two of \secref{sec:designspace} in detail, those that follow when the timer rule T1 or the renewal qualifier of P2 is weakened, and a walkthrough of the one that follows from the pseudocode of \cite{ref1}.  The two executions of \secref{sec:designspace} come from directed searches whose state constraints only prune behaviors, so each is a behavior of the full specification.

\noindent\textbf{Renewal across an acceptor restart.}  With three acceptors, a renewal and a single acceptor restart give two owners even with P2's renewal qualifier in place.  Acceptor $a_2$ promises ballot~2 to $p_2$, then crashes, forgetting the promise, and restarts into the full quarantine.  Proposer $p_1$, under the lower ballot~1, collects promises from $a_1$ and $a_3$, and $a_1$ then promises ballot~2 as well, which completes $p_2$'s prepare quorum with $a_2$'s forgotten promise.  Acceptor $a_3$ accepts $p_1$'s lease, and $p_2$'s \code{Accept} requests reach $a_1$ and $a_3$; at $a_3$ the higher ballot overwrites $p_1$'s accepted lease.  Proposer $p_2$ becomes active and renews under ballot~3, and $a_1$ and $a_3$ both report the exact lease being renewed, so the qualifier counts them as open.  When $a_2$ leaves quarantine it holds no promise, and it accepts $p_1$'s delayed \code{Accept} request.  Together with $a_3$'s earlier acceptance this completes $p_1$'s majority, and $p_2$'s renewal completes as well.  The quarantine outlasted $p_2$'s first attempt, but that attempt had already completed, on a majority that included an overwritten accepted lease.  TLC finds the execution in a 39-state trace (\code{PaxosLeaseRenewCrashSearch.cfg}), and the same search with A2's rule exhausts its space with no violation.  The reference model replays it (\code{renewal-over-overwritten-grant}).

\smallskip
\noindent\textbf{Early release across an acceptor restart.}  Proposer $p_2$ acquires through $a_1$ and $a_2$ under ballot~2, above $p_1$'s ballot~1, and releases at once, with its \code{Accept} request and its \code{Release} to $a_3$ both still in flight.  Acceptor $a_2$ crashes, forgetting its promise and its accepted lease, restarts and serves the full quarantine, and $p_1$ then acquires under ballot~1 through $a_2$ and $a_3$.  The stale \code{Accept} request reaches $a_3$ and, since ballot order permits it, reinstalls $p_2$'s released instance over $p_1$'s live lease; the stale \code{Release} then matches that instance exactly and clears it, together with its exclusion deadline.  A fresh attempt by $p_2$ under ballot~3 finds no lease at $a_1$, whose accepted lease it released, and none at $a_3$, and its \code{Accept} request overwrites $p_1$'s live accepted lease at $a_2$, so $p_2$ acquires while $p_1$ is still active.  The renewal qualifier is not involved, since $p_2$'s last attempt is fresh and the acceptors it asks hold no accepted lease, and no quarantine length prevents the execution, since the stale messages can be delayed arbitrarily.  With A2's rule, $a_3$ refuses to replace $p_1$'s live lease, the stale \code{Release} matches nothing, and $p_1$'s lease blocks $p_2$.  TLC finds the execution in a 40-state trace (\code{PaxosLeaseReleaseCrashSearch.cfg}), and the same search with A2's rule exhausts its space with no violation.  The reference model reproduces the execution and its repair (\code{release-after-stale-accept}).  Neither Keyspace nor ScalienDB implements early release, so this execution cannot occur in them.

\noindent\textbf{Timer placement (T1).}  T1 starts the attempt deadline when \code{Prepare} is sent and stores it as state.  Acceptor state is volatile, so evidence gathered before a crash must expire at a fixed time after it was created.  T1 assigns $t_d$ at P1 and checks it at P3 and P4, so the whole attempt, its unprocessed promises included, expires with it.

The pseudocode of \cite{ref1} starts the timer at prepare-quorum receipt instead.  The attempt then has no timer until the quorum arrives, a prepare quorum stays usable for an unbounded time, and no finite quarantine covers an attempt that has no deadline.  With overwriting acceptors, TLC finds the two-owner execution at the full quarantine bound in 27 states with two acceptors, and in 25 states with three acceptors and a single crashed acceptor, the one in the intersection of the two proposers' quorums.  The walkthrough below gives the failure in implementation terms.  In it, each acceptor replaces a live lease of the other proposer, and with A2's rule the two-acceptor configuration passes over 23{,}777{,}320 distinct states.

Keyspace and ScalienDB bound the attempt with a retry timeout instead, whose deadline exists only as dispatch order: a scheduling pause runs an expired attempt's handlers before the timeout fires, and TLC finds the two-owner execution under the shipped constants (\secref{sec:audit}, third finding).  Durable promises and Flease's clock-derived ballots \cite{ref10} also prevent the execution; the repairs are compared after the walkthrough.

\smallskip
\noindent\textbf{The 2012 pseudocode's timer rule.}  Figure~2 of \cite{ref1} labels ``start timer'' twice: in its row of event labels before the \code{Prepare} requests leave, and at a marker on the proposer's lane after the prepare responses arrive.  The pseudocode starts it in step~3, \code{Proposer::OnPrepareResponse}, when a quorum of empty \code{Prepare} responses has arrived, immediately before sending \code{Propose} requests.  The proof's prose says that the proposer starts its timer before sending \code{Propose} requests, which is consistent with both placements.

Under the pseudocode rule, an attempt has no deadline between sending \code{Prepare} and processing its prepare quorum.  The original restart quarantine of the maximal lease time $M$ outlasts forgotten exclusions and attempts whose timers have started.  It cannot outlast an attempt whose timer may start arbitrarily late.  No finite quarantine repairs that rule.

\code{tla/counterexamples/LateTimer.tla} changes only timer placement: \code{SendAccept} sets the pending deadline instead of \code{StartAcquire}.  With two proposers, two acceptors that overwrite, as in \cite{ref1}, and $D_P = D_A = D_Q = 2$, TLC finds a 27-state violation of \code{LeaseExclusivity}.  The execution proceeds as follows.

\begin{enumerate}
\itemsep0.15em
\item At $t=0$, $p_1$ and $p_2$ send \code{Prepare} with ballots $1$ and $2$.  Both acceptors promise in ballot order, and both proposers receive empty promise quorums.  Neither proposer has started its timer.
\item Both acceptors crash, forgetting their promises.  They restart and remain in quarantine until $t=2$.
\item At $t=1$, both proposers process their prepare quorums.  Each starts its timer with deadline $t=3$ and sends \code{Accept}.
\item At $t=2$, quarantine ends.  The acceptors accept ballot~$1$, then overwrite it with ballot~$2$; the crash erased the promises that would have rejected ballot~$1$.
\item Both proposers receive accepted quorums and activate.  At $t=2$, both own the lease until $t=3$.
\end{enumerate}

Every acceptor served its full quarantine.  The execution needs no message loss or clock error.  The original invariance argument omits acceptor state loss; its restart wait assumes that every attempt using forgotten state already has a bounded lifetime.

In implementation terms, ordinary scheduling delay separates receipt of the last \code{Prepare} response from sending the \code{Propose} requests.  A garbage-collection pause or a preempted virtual machine, together with acceptor crash and restart, supplies the whole schedule.  The proposer resumes with responses that remain acceptable under the pseudocode despite the loss of the promises they report.

A2's acceptor-side rule prevents this, since in step~4 each acceptor replaces a live lease of the other proposer.  Rule T1 prevents it as well, by storing the deadline when \code{Prepare} is sent and checking it before sending accepts and activating, so that unprocessed promises expire with the attempt.  With overwriting acceptors, T1 and a quarantine of $D_P$ are still not enough in general (the renewal execution of \secref{sec:designspace}); with A2's rule they are (\secref{sec:safety}).  The \code{late-promise-reuse} Python witness reproduces the unsafe schedule; its paired test uses the prepare-time rule, rejects the expired promise quorums, and sends no \code{Accept}.

Durable promises also prevent the execution, at the cost of a synchronous disk write on the prepare path.  An expiry token on \code{Prepare} responses applies the same time bound as T1.  Flease's clock-derived ballots use stronger clock assumptions \cite{ref10}.  A separate retry deadline can bound the attempt if stored and checked in every response handler.  The audited implementations use a scheduled retry callback without those checks; delayed dispatch defeats it (\ref{app:audit}).  For a separate retry bound $R$, quarantine must cover both the retry interval and the authority interval.  The implementation projection checks $D_Q \geq \max(D_P,R)$ under timely dispatch.  Storing and enforcing the retry deadline removes that dependence on callback ordering.

With three acceptors and majority quorums of two, TLC finds a 25-state two-owner trace under proposer and acceptor symmetry reduction.  Only the acceptor at the quorum intersection crashes; it forgets its promise, serves the full quarantine, and participates in both accept quorums.  The failure therefore also occurs when one acceptor remains available to each proposer throughout.  Separately, quarantine one unit below $D_P$ yields a 24-state trace with three acceptors.  The repository records these searches, \code{make paper-claims} checks that each recorded trace has the cited length, and \code{make paper-evidence-full} reruns them.

\smallskip
\noindent\textbf{Renewal check (P2).}  A report of the proposer's own lease is open only during a renewal of that exact lease.  The reported ballot must equal the renewal base captured at P1; ownership alone does not qualify a response.

A proposer abandons an attempt (P5) with \code{Accept} requests in flight and retries under a fresh ballot, and in between a competitor on its first attempt acquires under a lower ballot.  With overwriting acceptors, the late \code{Accept} requests overwrite the competitor's accepted lease at every acceptor, since ballot order permits it and overwriting revokes nothing, and the retrying proposer's prepare then finds its own lease everywhere it looks.  If ownership alone makes those responses open, it completes the acquisition while the competitor is still active.  TLC finds the execution in 36 states at the full quarantine bound with retry enabled (variant \code{StaleOwnerOpen.tla}), and no quarantine length prevents it, because the stale \code{Accept} request can be delayed arbitrarily.  Both audited implementations ship the unqualified rule (\secref{sec:audit}, seventh finding).  With A2's rule the stale \code{Accept} request cannot replace the competitor's accepted lease, and the same configuration with the qualifier dropped passes over 2{,}765{,}270{,}158 distinct states.

\section{Audit of Keyspace and ScalienDB}\label{app:audit}

This appendix gives the eight findings of \secref{sec:audit} in full, and the source-level evidence behind them.  The audit reads Keyspace at commit \code{a99f24a8} (2011) and ScalienDB at commit \code{60978146} (2013), the final commits of both repositories.

\subsection{Findings}

\begin{enumerate}\itemsep0.25em
\item \textbf{Both systems implement the unsafe timer rule of the pseudocode, but an unrelated retry timeout accidentally prevents the late-timer executions.}  The rule that shipped is the \emph{unsafe} one of the pseudocode, not the prepare-time reading of the figure: \code{StartProposing()} sets \code{expireTime = Now() + duration}, so the authority clock starts when the prepare quorum is in hand, the rule that \ref{app:latetimer} shows unsafe with overwriting acceptors, which both systems have.  What both contain instead is a mechanism \cite{ref1} never mentions, an attempt-restart timeout of $R = 2$ seconds armed when \code{Prepare} is broadcast, whose firing re-prepares under a fresh proposal identifier and thereby discards the abandoned attempt's responses.  This is T1's deadline enforced by control flow rather than stored as data, which \ref{app:latetimer} shows insufficient.  Neither codebase presents it as a safety mechanism; it serves as one by accident.
\item \textbf{That accidental prevention works only under certain conditions, which the implementations accidentally satisfy.}  We built a checked projection of the implementations onto their lease acquisition state machine (\code{tla/spec/PaxosLeaseImpl.tla}) and checked when the accidental repair works, in acquisition configurations without renewal.  Under timely timeout dispatch the restart condition for the late-timer executions is
\[
  D_Q \geq \max(D_P, R),
\]
checked at its boundary and refuted one unit below the $R$-dominant boundary and one unit below the common boundary $D_P = R$; the boundary run also refutes the additive hypothesis $D_Q \geq R + D_P$.  The shipped ordering of the constants passes exhaustively in these configurations; the seventh finding shows the shipped protocol unsafe for another reason.
\item \textbf{The event loop's handler ordering can produce a two-owner execution.}  Neither proposer stores when its attempt began; the $R$ bound holds only because each event-loop iteration runs due timers before socket events.  The real assumption is therefore stronger than freedom from suspension: no message handler belonging to an expired attempt ever executes past the deadline before the timeout transition runs, and the loop structure does not enforce that: the poll blocks until the next timer is due, so a socket that becomes ready just before the deadline has its handlers, and any burst behind them, dispatched before the next timer scan.  A process suspension beginning after a timer scan and ending inside the poll is one instance; ordinary event-loop overrun is another.  TLC finds the resulting two-owner execution with the shipped constants in 27 states with two independently crashing acceptors, and in 25 states in the colocation-valid configuration described below, where only the quorum-intersection acceptor crashes; the \code{suspended-event-loop} witness replays it at \code{IsLeaseOwner()} level.  The shipped constants are therefore not enough to call the implementations safe.
\item \textbf{Quorums are counted by message rather than by acceptor identity, which is safe only because their TCP transport does not deliver a response twice.}  Both systems count quorum responses without acceptor identity, so one response delivered twice counts as two votes.  We found no path that retransmits a response: pending writes are discarded on disconnect, and the inspected TCP connections supply non-duplicating delivery within a connection.  The scalar counting therefore rests on an application-level at-most-once assumption that is plausible for these implementations but stated and checked nowhere, and the weakened variant \code{ScalarQuorumCounting.tla} shows what it protects: under a transport that may re-deliver, TLC finds a two-owner execution in 18 states.
\item \textbf{Every lease timer reads the wall clock, and neither system defends both of the directions in which a wall clock is unsafe.}  Clock errors are not symmetric between the roles: backward or slow steps are unsafe on a proposer (authority extends) and merely prolong exclusion on an acceptor, while forward or fast steps are unsafe on an acceptor (exclusion shrinks) and harmless on a proposer.  Keyspace uses \code{gettimeofday()} raw and is exposed in both unsafe directions; ScalienDB repairs backward steps well, through a correction thread, but lets forward steps pass uncorrected.  Neither system calls the monotonic clocks its platforms provided, \code{clock\_gettime(CLOCK\_MONOTONIC)} on Unix and \code{GetTickCount64} or \code{QueryPerformanceCounter} on Windows.
\item \textbf{The learner is where ownership becomes visible, and its remote expiry rests on an undocumented delivery bound.}  Ownership is application-visible only through the learner, and the learner restarts the countdown from the moment the message arrives rather than from the moment the lease began.  In both systems the owner's own learner installs the absolute expiry the proposer computed, so the two-owner executions above carry through to two simultaneously application-visible masters; but every \emph{remote} learner sets its expiry to arrival time plus the remaining duration minus a 500 msec hedge, so the hedge is conservative only while \code{LearnChosen} delivery takes less than 500 milliseconds, another undocumented timing assumption.
\item \textbf{A proposer treats any lease it owns as an open response even when it is no longer the active owner.}  Both proposers treat a reported lease they own as an open response whether or not they are currently the active owner: \code{StartProposing} proceeds, with a full fresh duration, whenever the discovered lease owner is the node itself.  \ref{app:latetimer} shows the rule unsafe with abandon and retry (P5), of which the retry timeout of the first finding is an instance: an \code{Accept} request from an abandoned attempt can overwrite a competitor's live accepted lease and then serve as its sender's own justification, and TLC finds the resulting two-owner execution in 36 states at the full quarantine bound, with no dispatch delay and no clock error.  The execution does not depend on connectionless delivery: under the connection-oriented refinement, in which a crash discards the messages addressed to the crashed process, TLC finds it at the same depth, with the stale \code{Accept} requests sent after the acceptors restart, as a reconnecting writer does.  The renewal qualifier that P2 states prevents this execution: an own lease is an open response only for a renewal attempt, and only for the exact lease captured as its base while it is still held.  A2's refusal prevents it as well.  The composition is checked in the implementation projection itself, not only in the base model: with three ballots, the shipped timely constants, and the connection-lifecycle transport, the projection combines the shipped retry timeout with the shipped own-lease rule, and TLC finds a 35-state two-owner execution (\code{PaxosLeaseImplStaleOwner.cfg}).  Timer dispatch is timely throughout that execution: it needs no suspension, no dispatch delay, and no clock error.  These traces crash both acceptors of a two-acceptor system, so they model acceptors that crash independently of the proposers; whether they arise in a two-node deployment, where each node hosts both roles, is not established.
\item \textbf{The activation guard reads the clock twice, and an unsigned subtraction can wrap so that a proposer activates a lease that has already expired.}  The Phase~2 handlers read the clock once for the expiry guard and again for the activation-margin subtraction, on \code{uint64\_t} deadlines.  If the clock crosses the deadline between the two reads, through a pause between the calls or a forward step, the unsigned subtraction wraps, the margin check passes, and the proposer activates an expired lease; Keyspace additionally broadcasts the wrapped value as the lease duration.  The window is narrow, but it is a time-of-check to time-of-use race on the safety-critical comparison, and the repair, one clock read per handler with a compare-before-subtract, is standard.
\end{enumerate}

\subsection{Timer Rule and Retry Timeout}

In both systems \code{StartProposing()} sets \code{expireTime = Now() + duration} (Keyspace \code{PLeaseProposer.cpp}, ScalienDB \code{PaxosLeaseProposer.cpp}).  The attempt-restart timeout is \code{ACQUIRELEASE\_TIMEOUT}, 2 seconds in both, armed by \code{StartPreparing()}; when it fires, \code{OnAcquireLeaseTimeout()} calls \code{StartPreparing()} again under a fresh proposal identifier.  The maximal lease time is \code{MAX\_LEASE\_TIME}, 7 seconds in Keyspace and 3 in ScalienDB, and both the lease duration and the startup quarantine are set to it.

\code{OnProposeResponse} contains the unsigned-arithmetic race of the eighth finding.  It tests \code{state.expireTime < Now()} and returns if the lease has expired, then, with a separate clock read, tests \code{state.expireTime - Now() > 500} before activating; \code{expireTime} is \code{uint64\_t} (\code{PLeaseState.h}).  A clock that crosses the deadline between the two reads wraps the subtraction to nearly $2^{64}$, which passes the margin test, and Keyspace then passes \code{state.expireTime - Now()}, a third read, to \code{LearnChosen} as the lease duration.  The \code{unsigned-expiry-underflow} witness records the arithmetic.

\subsection{TLA+ Model of the Shipped Protocol}

\code{tla/spec/PaxosLeaseImpl.tla} has overwriting acceptors, as both implementations do, and differs from the base specification with overwriting acceptors in three audited rules: \code{pendingDeadline} is set in \code{SendAccept} rather than \code{StartAcquire}; an attempt deadline of $R$ units, set at \code{StartAcquire}, abandons the ballot when it fires; and a reported own lease counts as open whenever the proposer owns it, with no renewal base.  It keeps three behaviors of the base specification that differ from the shipped code: its acceptors reject lower ballots after an accept, its proposer counts only responses that report no lease or its own lease where the shipped proposer counts every non-rejecting response and follows the highest-ballot report, and a \code{LearnChosen} naming another owner does not stop the local proposer.  Its passing runs therefore do not carry over to the shipped code, and the delayed-dispatch schedule below also requires the \code{LearnChosen} messages between the two owners' nodes to stay undelivered.  Unlike the shipped counter-major identifiers, whose attempt counter restarts, the projection keeps a restarted proposer's attempt counter, so its ballots never decrease across a restart, and its passing configurations do not cover ballot regression after a proposer restart.  It has no release, as the shipped systems have none, so the incarnation execution of \secref{sec:safety} does not arise in it.  The Phase~2 delivery action has no attempt-deadline guard, matching \code{OnProposeResponse}.  The constant \code{TimelyDispatch} selects the timeout semantics.  Under timely dispatch, advancing time abandons every overdue attempt before anything else happens, which assumes that overdue attempts are abandoned before any further response handler executes.  Under delayed dispatch, expiry only enables an abandonment action that any amount of message processing may postpone, which covers both a suspension and ordinary handler overrun.

\ref{app:artifact} lists the module's timer and quarantine configurations, in $(D_P, R, D_Q)$, with their state counts.  Since the boundary passes and one unit below it fails, the condition the shipped constants must satisfy in these configurations is $R \leq M$, not $R < M$, and both implementations satisfy it strictly.  The coverage lemma behind the bound is the TLAPS obligation \code{ImplQuarantineCoversForgottenFacts} (\ref{app:invariants}).

The two-acceptor delayed-dispatch trace crashes both acceptors, and in the deployed systems a node crash restarts the colocated proposer and bumps the durable restart counter in its proposal identifiers, so in a two-node deployment that trace would change both proposers' ballot ordering.  \code{PaxosLeaseImplDelayedColocated.cfg} therefore uses three acceptors and restricts \code{CrashableAcceptors} to the single acceptor in the intersection of the two proposers' quorums.  No proposer-hosting node then crashes, no restart counter moves, and the node-identity tiebreak between equal attempt counters is legitimate, so its 25-state trace is valid for the colocated deployment without modeling restart counters; the schedule below is of the same kind: only the quorum-intersection acceptor crashes, and the higher-ballot proposer's handlers run after its retry deadline.

\subsection{A Two-Owner Schedule under Delayed Dispatch}

A process that merely resumes after a long pause fires the overdue retry at the next timer scan and discards the stale responses, so the schedule pauses inside one loop iteration, between \code{RunTimers()} and the return of \code{IOProcessor::Poll()} (\secref{sec:audit}, third finding).  It crashes exactly one node, the acceptor in the intersection of the two quorums, which hosts neither active proposer.

\begin{enumerate}
\itemsep0.15em
\item Node 2's proposer broadcasts \code{Prepare}, and acceptors 1 and 2 promise.
\item Node 1 alone crashes after promising, forgets the promise, restarts, and begins its full quarantine.
\item Node 2 processes its prepare quorum and starts proposing: its authority interval starts now, and its \code{Accept} requests are delayed.  Its loop then performs the last timer scan before the retry deadline and blocks in the poll.  The suspension begins there and outlasts node 1's quarantine.
\item Node 0's proposer completes an entire fresh acquisition against acceptors 0 and 1, and its learner installs the absolute expiry: node 0 is the application-visible master.
\item The stale \code{Accept} requests are accepted at node 1, overwriting node 0's accepted lease without revoking node 0's authority, and, on resume, at node 2's own acceptor.  Node 2's stale ballot exceeds node 0's fresh one because the attempt counters are equal, neither node restarted, and the tie breaks on node identity, exactly the counter-major layout.
\item Node 2's queued responses are then dispatched before the overdue timeout callback, the handler finds the lease expiry unreached with more than its 500 msec margin remaining, the proposer activates, and its own learner installs the absolute expiry.  Nodes 0 and 2 are now simultaneously application-visible masters.
\end{enumerate}

\code{python/paxoslease/event\_loop.py} replays this schedule against an executable abstraction of the shipped node, with proposer, acceptor, and learner colocated, as the \code{suspended-event-loop} witness, asserted at \code{IsLeaseOwner()} level; a paired run shows the same schedule refused without the suspension.  Exceeding the 500 msec margin requires only a virtual-machine pause, a swap stall, a \code{SIGSTOP}, or a long enough handler burst.

\subsection{Quorum Counting}

Keyspace's proposer counts scalars (\code{numReceived++}, \code{numAccepted++}), and ScalienDB's vote object takes a node identifier but uses it only for a membership test before incrementing a scalar (\code{MajorityQuorum.cpp}).  Pending writes are discarded on disconnect in \code{TCPConn::Close} (Keyspace) and \code{TCPConnection::Close} (ScalienDB).  A proof of at-most-once delivery across connection replacement would also have to exclude simultaneous old and new connections to one peer, regenerated responses on reconnect, and handler re-entry, which we have not done, and Keyspace's tree contains an unused UDP transport.  The implementation contract states the rule disjunctively: count by identity, or document and preserve at-most-once delivery per response.

\begin{table*}[!t]
\centering
\scriptsize
\setlength{\tabcolsep}{4pt}
\renewcommand{\arraystretch}{1.0}
\begin{tabular*}{\textwidth}{@{\extracolsep{\fill}}lllll}
\toprule
\textbf{Configuration} & \textbf{$(D_P, D_A, D_Q)$} & \textbf{Purpose} & \textbf{Distinct states} & \textbf{Result} \\
\midrule[\heavyrulewidth]
\multicolumn{5}{l}{\emph{The specification (\code{PaxosLeaseChecked.tla}): A2's refusal, owners compared as incarnations}} \\
Base & $(2,2,2)$ & acquisition races, two proposers, three acceptors & 491,037 & pass \\
Renew/release & $(2,2,2)$ & renewal and exact-instance release, one proposer & 7,717 & pass \\
Crash/restart & $(2,2,2)$ & two proposers race across crash and restart & 37,425,056 & pass \\
Drift & $(1,2,2)$ & $D_P < D_A$, one proposer; clock-rate error is not modeled & 40,784 & pass \\
Quarantine $=D_P$ & $(1,2,1)$ & boundary: quarantine below acceptor exclusion & 76,749,192 & pass \\
Quarantine $=D_P$ & $(2,3,2)$ & boundary at a second duration pair & 36,772,096 & pass \\
Redeliver & $(2,2,2)$ & redelivering transport, identity counting, no crash & 1,857,563 & pass \\
Release race & $(2,2,2)$ & release, two proposers, three acceptors & 23,743,497 & pass \\
Renew+Crash, 3 acceptors$^\dagger$ & $(2,2,2)$ & renewal across crash and restart, symmetry reduced & 3,863,685,436 & stopped, no violation \\
Release+Crash, 3 acceptors$^\dagger$ & $(1,1,1)$ & release across crash and restart, symmetry reduced & 4,461,465,229 & stopped, no violation \\
Retry$^\dagger$ & $(2,2,2)$ & explicit abandon and retry, three ballots, crash & 2,179,760,458 & pass \\
Renew+Retry$^\dagger$ & $(2,2,2)$ & renewal and abandon together, crash & 2,411,951,538 & pass \\
Retry+Redeliver$^\dagger$ & $(2,2,2)$ & abandoned attempts under redelivery & 53,723,103 & pass \\
Renew+Release+Retry$^\dagger$ & $(2,2,2)$ & stale Phase~2 traffic around renewal and release, one proposer & 204,050 & pass \\
Incarnation (directed)$^\dagger$ & $(2,2,2)$ & the search of the \code{NodeOwner} execution & 220,119 & pass \\
Renew+Crash (directed)$^\dagger$ & $(2,2,2)$ & the search of the renewal execution & 6,330,050 & pass \\
Release+Crash (directed)$^\dagger$ & $(1,1,1)$ & the search of the early-release execution & 224,164,193 & pass \\
Lease+Paxos & n/a & abstract composition barrier & 2,095 & pass \\
\midrule
\multicolumn{5}{l}{\emph{The specification one unit below the quarantine bound (\code{PaxosLease.tla})}} \\
Unsafe quarantine & $(1,2,0)$ & quarantine one unit below $D_P$ & not exhaustive & violated (25-state trace) \\
Unsafe quarantine & $(2,3,1)$ & quarantine one unit below $D_P$, $D_A$ larger & not exhaustive & violated (26-state trace) \\
\midrule
\multicolumn{5}{l}{\emph{Overwriting acceptors, the rule of \cite{ref1} (\code{RefuseLiveOverwrite = FALSE})}} \\
Renew+Crash (directed)$^\dagger$ & $(2,2,2)$ & renewal across one restart, three acceptors & not exhaustive & violated (39-state trace) \\
Release+Crash (directed)$^\dagger$ & $(1,1,1)$ & early release across one restart & not exhaustive & violated (40-state trace) \\
\bottomrule
\end{tabular*}
\caption{\label{tab:configs}TLC checks of the specification.  Passing counts are exhaustive for the finite configurations, except that Retry and Renew+Retry are exhaustive only up to TLC's fingerprint-collision estimate (\ref{app:artifact}), subject to the per-configuration bounds on in-flight messages, on proposer restarts, and on the time horizon, which is at most two duration units in every configuration of the specification except the renewal-and-release ones, which run to three; the directed rows exhaust only the space their state constraint leaves, which only prunes behaviors.  Violated runs stop at the counterexample; their trace lengths are minimum-depth under breadth-first search and stable across runs, while the number of states a parallel search happens to explore before finding the violation is not, so it is not cited.  The two boundary rows establish $D_Q \geq D_P$ as the tight bound in the checked model together with the two rows one unit below it.  The directed rows of the specification extend \code{PaxosLease.tla} directly, under the same assumptions as \code{PaxosLeaseChecked.tla}.  The two rows marked stopped give the distinct states explored before the search was stopped, and are not exhaustive.  Rows marked $\dagger$ are recorded runs, checked by \code{make paper-claims} and re-run by \code{make paper-evidence-full}, except the two stopped rows, whose counts are the last progress reports in their recorded logs; the others run in \code{make paper-evidence}.}
\end{table*}

\begin{table*}[!t]
\centering
\scriptsize
\setlength{\tabcolsep}{3pt}
\renewcommand{\arraystretch}{1.0}
\begin{tabularx}{\textwidth}{l>{\raggedright\arraybackslash}X>{\raggedright\arraybackslash}X>{\raggedright\arraybackslash}X>{\raggedright\arraybackslash}X}
\toprule
\textbf{Rule} & \textbf{Base module} & \textbf{Variant/projection} & \textbf{Python witness} & \textbf{Source audit} \\
\midrule[\heavyrulewidth]
Proposer lease timeout (P1, T1) & \code{StartAcquire} stores $t_d$ & \code{LateTimer} (overwriting) & simulator, timer witnesses & unsafe rule shipped \\
Retry time bound (P5) & \code{AbandonAttempt}, Retry cfg & impl projection ($R$) & event-loop witness & undocumented $R \leq M$ \\
Exact-instance release (A3) & \code{DeliverRelease} & \code{OwnerOnlyRelease} & release witness & no release path shipped \\
Restart quarantine (A4, T3) & \code{RestartAcceptor} & unsafe-quarantine cfgs & quarantine witness & full-$M$ quarantine \\
Identity counting (P2) & response sets, Redeliver & \code{ScalarQuorumCounting} & duplicate witness & scalar counters, TCP accident \\
Own-lease-open qualifier (P2) & \code{DeliverPromise} & \code{StaleOwnerOpen} (overwriting) & stale-owner-open witness & unqualified rule shipped \\
Acceptor-side refusal (A2) & \code{DeliverAcceptReq}, \code{SameOwner} & renewal and release searches, \code{NodeOwner} & renewal, release and incarnation witnesses & overwriting rule shipped \\
Lower ballots after accept (A2) & \code{DeliverAcceptReq} & n/a (argued in A2) & n/a & not in the shipped acceptors \\
Learner deadlines (L1) & not modeled & n/a & n/a & remote re-anchoring \\
Clock discipline & not modeled & n/a & clock-source, underflow witnesses & wall clock, double clock read \\
Leader lifecycle (M1--M5) & composition model & n/a & \code{leased\_paxos.py}, tests & per-slot promises \\
\bottomrule
\end{tabularx}
\caption{\label{tab:evidence}Each rule of \secref{sec:algorithm} and~\ref{sec:paxos} mapped to the models and tests that check it.  ``Not modeled'' marks the implementation-layer rules that the base transition system deliberately abstracts away; their evidence is executable witnesses and the audit of \ref{app:audit}.}
\end{table*}

\subsection{Learners}

Application code reads \code{IsLeaseOwner()} in \code{PLeaseLearner.cpp} and \code{PaxosLeaseLearner.cpp}, which the \code{LearnChosen} broadcast populates at every node, the sender included.  The owner's learner installs \code{msg.localExpireTime}, the absolute expiry the proposer computed; every other learner installs \code{Now() + msg.duration - 500}.  The source comments call the subtraction a conservative estimate, which it is while delivery takes less than 500 milliseconds; a \code{LearnChosen} delayed by $\Delta > 500\,\mathrm{ms}$ leaves every remote learner believing in the owner's authority $\Delta - 500\,\mathrm{ms}$ past its real end.  Remote learners do not act as owner, but master lookups and election suppression read this state.

\subsection{Clock Sources}

In Keyspace, \code{Now()} is \code{GetMilliTimestamp()}, which is \code{gettimeofday()} raw, with no clamp and no correction (\code{Platform.cpp}; the \code{unsafe-clock-source} witness).  In ScalienDB, every lease-layer call reaches the \code{Now()} of \code{Time.cpp}: \code{gettimeofday()} plus a global correction offset, with a per-thread monotonic clamp.  A clock thread started unconditionally at boot (\code{Main.cpp}) samples the clock every 5 milliseconds and, on a backward step, permanently raises the offset so that corrected time resumes at its previous value plus one resolution step.  The correction is best effort: it loses any time that passes while the sampling thread is delayed, and it gives no bound on the rate of the corrected clock.  Forward steps pass through uncorrected in both systems.  ScalienDB's Windows path in \code{Time.cpp} builds an elapsed-time scheme on \code{timeGetTime()}, milliseconds since boot, and never runs it: the enclosing \code{gettimeofday()} opens with \code{return gettimeofday\_win(tv, NULL);} and everything below that return is unreachable.

\subsection{Connection-Oriented Transport}

Under \code{CrashDropsIncoming}, messages addressed to a process are discarded when it crashes and again when it restarts, so every delivered message was sent after its addressee's most recent restart; messages a crashed process already sent survive.  The verdicts do not change.  The late-timer execution is unchanged at 27 states, because its stale messages are promises, sent by the acceptors before they crash.  TLC finds the stale-owner execution at 36 states as well: the acceptors crash and restart first, and the abandoned attempt's \code{Accept} requests are sent afterward, on fresh connections, as a reconnecting writer does.  In the implementation projection, the colocation-valid configuration under this transport (\code{PaxosLeaseImplDelayedColocatedTcp.cfg}) finds a 25-state two-owner execution, and the stale-owner projection of the seventh finding runs under the same transport.  Connection teardown masks none of these executions; the transport property safety rests on is the at-most-once delivery of the fourth finding.

In neither the renewal nor the early-release execution of \secref{sec:designspace} is a message that was sent before an acceptor's restart delivered to that acceptor after it, so both are also executions under this transport.

\section{Repository Inventory}\label{app:artifact}

This appendix inventories the evidence behind \secref{sec:checked}: every TLC configuration of the specification with its state count and verdict, the mapping from each rule to the model or test that checks it, the implementation-projection configurations, and the Python models.

\tabref{tab:configs} lists the configurations of \code{tla/spec/PaxosLease.tla}.  The two boundary rows are the separated-duration pairs of \secref{sec:safety}, and the rows that combine renewal, retry, release, redelivery and crash are the interaction configurations of \secref{sec:checked}.

Inside such a configuration TLC enumerates every reachable state and finds no violation of \code{LeaseExclusivity}; about larger ones it says nothing, and \secref{sec:limitations} says which of these bounds the result actually rests on.  \tabref{tab:configs} lists them all.  They cover the separated-duration boundary pairs of \secref{sec:safety}, explicit retry at the quarantine boundary, and the redelivering transport.  Further configurations test feature interactions, combining renewal, retry, release, redelivery, and crash, and two of them run release with two proposers.  Features that pass in isolation need not pass together, as the stale-owner execution shows, so these configurations combine them.  A further family of configurations projects the rules the two audited implementations actually ship, in particular a retry timeout in place of a stored attempt deadline, and reproduces both machine-found failures under the constants those systems ship (\secref{sec:audit}).

The five variants of \tabref{tab:variants} each weaken one rule.  Three of them weaken a rule that the protocol needs with A2's rule in place: the incarnation clause of A2 itself, in the 43-state execution of \secref{sec:safety}, and the two below.  If release matches on owner alone, then a \code{Release} message delayed past a re-acquisition by the same owner erases the newer lease.  The owner releases instance $(p_1, 1)$ and re-acquires as $(p_1, 2)$, the stale \code{Release} then clears the ballot-2 lease at the acceptors, and a second proposer acquires while $p_1$ is still active, in a 36-state trace.  With quorums counted by message under a transport that may re-deliver, the fourth finding of \secref{sec:audit}, a single acceptor's doubled response is a quorum, and two proposers become active in 18 states with no crash, no restart, and no clock movement at all.  The other two, \code{LateTimer} and \code{StaleOwnerOpen}, weaken T1 and P2 with overwriting acceptors, the rule of \cite{ref1} and of both audited implementations: the timer placement of \ref{app:latetimer}, which walks its 27-state trace through in full, and the stale-owner execution, whose 36-state search is the multi-hour recorded run \code{make counterexamples-staleowner}.

The safety argument of \secref{sec:safety} rests at several points on inequalities between durations, and those inequalities are proved mechanically, by TLAPS, in \code{tla/proof/PaxosLeaseProof.tla}: the inequalities behind timer containment, the quarantine bound and its tightness, the monotonicity of the deadline cap used by the finite model, and the clock-rate condition of T2.  They are arithmetic facts; the protocol-level premises they are applied to are the argument of \secref{sec:safety}, checked by TLC.

The accompanying repository also contains \code{python/demo.py}, a dependency-free, single-file implementation of the algorithm of \secref{sec:algorithm}, written to be read by practitioners.  It runs three nodes in one event loop on real monotonic-clock timers, and each node hosts a proposer, an acceptor, and a learner.  Every method names the step it implements, P1 through P5, A1, A2 and A4, L1, and the timing rules, so the file reads against the paper.  Where this paper shows that a plausible alternative to a rule is unsafe, the comment at that rule says so and points at the section.  The program demonstrates normal operation: a node acquires the lease, renews it at half-life, and dies; a survivor takes over once lease and quarantine have run out; and a referee object receives every learner-level ownership claim and asserts throughout that no two nodes are ever owner at once.

Every configuration bounds the number of in-flight messages with the constant \code{MaxNetwork}, so a passing count is exhaustive for the constrained state graph (\secref{sec:limitations}).  In the configurations of \tabref{tab:configs}, the bound is 3 in the single-proposer configurations, 7 in the incarnation search, the directed release search and Release+Crash with three acceptors, and 4 in the others, so the three-acceptor passing rows admit fewer interleavings than the three-acceptor violations need.  TLC stores visited states as 64-bit fingerprints: for Retry, Renew+Retry and the stale-owner check with A2's rule, its own estimate of the probability that a fingerprint collision left a state unexplored is 0.2, 1.0 and 0.33, and 1.2, 1.5 and 2.0 by its optimistic calculation; for every other passing run it is below 0.003.  A violation found under the bound is a legal trace of the unbounded model, so the bound does not weaken the violated rows.  The violating executions of this paper need at most six concurrent messages, except early release (\secref{sec:designspace}) and the incarnation execution (\secref{sec:safety}), which reach seven.

Not every rule of \secref{sec:algorithm} and~\ref{sec:paxos} lives in the base module, and \tabref{tab:evidence} states which model or test checks which rule.  The proposer, acceptor and timing rules are in the base module, checked in every configuration.  The learner rule L1 and the clock discipline concern how an implementation surfaces activation and expiry, which the base module abstracts away, so their evidence is the executable witnesses and the source audit of \ref{app:audit}.  The leader lifecycle M1 through M5 is checked by the TLA+ composition model, which checks the admission barrier between the two layers, and by the Python model, which runs the recovery, repair and append rounds against ballot-checking durable acceptors.

The implementation projection \code{tla/spec/PaxosLeaseImpl.tla} of \secref{sec:audit} has seven timer and quarantine configurations, written $(D_P, R, D_Q)$, beside three transport and stale-owner projections.  Three pass under timely timeout dispatch: $(D_P, R, D_Q) = (2,2,2)$ over 37{,}160{,}904 distinct states, $(1,2,2)$ over 37{,}476{,}080, and the shipped ordering $(2,1,2)$ over 18{,}160{,}464.  Four violate \code{LeaseExclusivity}: quarantine one unit below the boundary, at $(2,2,1)$ and $(1,2,1)$, each in 26 states; the shipped constants under delayed dispatch with two acceptors, in 27 states; and the colocation-valid form with three acceptors, in which only the quorum-intersection acceptor may crash, in 25 states.  A further projection combines the shipped retry timeout with the unqualified own-lease rule under timely dispatch and the connection-lifecycle transport, and finds a 35-state two-owner execution (\secref{sec:audit}, seventh finding).

The Python layer holds two deterministic models with virtual time, a lease-only model and a composed lease-plus-Paxos model.  They encode the rules a second time, independently of the TLA+, and assert the safety invariants after every step of every schedule they run.  TLC searches all schedules within a configuration; the Python models replay specific ones.  Eighteen structured witnesses keep every known failure executable.  Seventeen of them must fail with their specific expected message, so an unrelated failure does not satisfy the test; the eighteenth demonstrates the fenced repair of an external resource.

The five mechanically generated variants of \tabref{tab:variants} each weaken one rule, and TLC finds an exclusivity violation in each; those counterexamples are the stronger class, since there the model checker finds the violating schedule itself.  TLAPS proves the six arithmetic obligations listed in \ref{app:invariants}.  The multi-hour searches, the rows of \tabref{tab:configs} marked $\dagger$, the stale-owner runs, the three-acceptor counterexamples and the colocation-valid search, are recorded runs; \code{make paper-claims} checks their recorded state counts and trace lengths against the numbers this paper cites, and \code{make paper-evidence-full} re-runs them, except the two stopped searches, which are recorded logs only.

\section{How This Paper Was Written}\label{sec:llm}

Large language models produced much of the TLA+ specifications and TLC configurations, the TLAPS proofs, the Python programs, the source audit, and some of the article prose, under the author's manual, iterative direction, with line-by-line reviews and rewrites; the overall process spanned months.  Make targets automated the checking: they run the TLA+ parser, the TLC model checker, the TLAPS proof manager, and the Python tests, and record the results that the paper cites.  A full verification run of the repository, including parsing, every TLC configuration and counterexample search, the TLAPS proofs, and the Python tests, takes more than 40 hours on a 4-core, 8-thread Xeon server with 64~GB of memory, most of it in the multi-billion-state searches.

Machine assistance and automation reduced the recurring work of maintaining the specification, configurations, counterexamples, Python models, and verification pipeline.  Revisions were cheap enough that rerunning the checks remained practical after each change, making re-checking a natural part of the workflow.  The source audit also benefited from assistance in tracing two old implementations against the model's rules.

Every generated formal artifact is independently checkable.  TLA+ parsing checks syntax, TLC checks invariants over each configured state graph, and TLAPS checks the arithmetic proof obligations.  Weakened variants must produce their expected counterexamples, and regression targets check that each violation witness fails for the advertised reason.


\begin{thebibliography}{15}
\itemsep0em
\bibitem{ref1} M. Trencs\'eni, A. Gazs\'o, and H. Reinhardt. ``PaxosLease: Diskless Paxos for Leases.'' arXiv:1209.4187, 2012. \url{https://arxiv.org/abs/1209.4187}
\bibitem{ref2} L. Lamport. ``Paxos Made Simple.'' \textit{ACM SIGACT News}, 32(4), 2001.
\bibitem{ref3} C. Gray and D. Cheriton. ``Leases: An Efficient Fault-Tolerant Mechanism for Distributed File Cache Consistency.'' \textit{SOSP}, 1989.
\bibitem{ref4} D. Ongaro and J. Ousterhout. ``In Search of an Understandable Consensus Algorithm.'' \textit{USENIX ATC}, 2014.
\bibitem{ref5} S. Chand, Y. Liu, and S. Stoller. ``Formal Verification of Multi-Paxos for Distributed Consensus.'' \textit{FM}, 2016.
\bibitem{ref6} M. Burrows. ``The Chubby Lock Service for Loosely-Coupled Distributed Systems.'' \textit{OSDI}, 2006.
\bibitem{ref7} T. Chandra, R. Griesemer, and J. Redstone. ``Paxos Made Live: An Engineering Perspective.'' \textit{PODC}, 2007.
\bibitem{ref8} F. Hupfeld, B. Kolbeck, J. Stender, M. H\"ogqvist, T. Cortes, J. Mart\'i, and J. Malo. ``FaTLease: Scalable Fault-Tolerant Lease Negotiation with Paxos.'' \textit{HPDC}, 2008.
\bibitem{ref9} L. Lamport. ``The Part-Time Parliament.'' \textit{ACM Transactions on Computer Systems}, 16(2), 1998.
\bibitem{ref10} B. Kolbeck, M. H\"ogqvist, J. Stender, and F. Hupfeld. ``Flease: Lease Coordination without a Lock Server.'' \textit{IPDPS}, 2011.
\bibitem{ref11} U. Sharma, R. Jung, J. Tassarotti, M. F. Kaashoek, and N. Zeldovich. ``Grove: a Separation-Logic Library for Verifying Distributed Systems.'' \textit{SOSP}, 2023.
\bibitem{ref12} I. Moraru, D. G. Andersen, and M. Kaminsky. ``Paxos Quorum Leases: Fast Reads Without Sacrificing Writes.'' \textit{SoCC}, 2014.
\bibitem{ref13} H. Cirstea, M. A. Kuppe, B. Loillier, and S. Merz. ``Validating Traces of Distributed Programs Against TLA+ Specifications.'' arXiv:2404.16075, 2024.
\bibitem{ref14} M. Trencs\'eni and A. Gazs\'o. ``Keyspace: A Consistently Replicated, Highly-Available Key-Value Store.'' arXiv:1209.3913, 2012. \url{https://arxiv.org/abs/1209.3913}

\bibitem{ref15} M. Trencs\'eni and A. Gazs\'o. ``ScalienDB: Designing and Implementing a Distributed Database using Paxos.'' arXiv:1302.3860, 2013. \url{https://arxiv.org/abs/1302.3860}
\end{thebibliography}
\end{document}